\documentclass{bmcart}

\usepackage{amsthm,amsmath}
\usepackage[utf8]{inputenc} 
\usepackage{graphicx}
\usepackage{dcolumn}
\usepackage{bm}
\usepackage{textcomp}
\usepackage{revsymb}

\startlocaldefs
\endlocaldefs

\begin{document}

\begin{frontmatter}

\begin{fmbox}
\dochead{Research}


\title{Evidence for kinetic limitations as a controlling factor of Ge pyramid formation: \\A study of structural features of Ge/Si(001) wetting layer formed by Ge deposition at room temperature followed by annealing at 600$^{\,\circ}$C}


\author[
   addressref={aff1},                   
   email={storozhevykh@kapella.gpi.ru}   
]{\inits{MS}\fnm{Mikhail S} \snm{Storozhevykh}}
\author[
   addressref={aff1},
   email={arapkina@kapella.gpi.ru}
]{\inits{LV}\fnm{Larisa V} \snm{Arapkina}}
\author[
   addressref={aff1,aff2},
	corref={aff1},                       
   email={vyuryev@kapella.gpi.ru}
]{\inits{VA}\fnm{Vladimir A} \snm{Yuryev}}


\address[id=aff1]{
  \orgname{A.\,M.\,Prokhorov General Physics Institute of the Russian Academy of Sciences, 38 Vavilov Street},                     %
  \postcode{119991}                                
  \city{ Moscow},                              
  \cny{Russia}                                    
}

\address[id=aff2]{
  \orgname{Technopark of GPI RAS, 38 Vavilov Street},                     %
  \postcode{119991}                                
  \city{ Moscow},                              
  \cny{Russia}                                    
}
%

\begin{artnotes}
\end{artnotes}

\end{fmbox}


\begin{abstractbox}

\begin{abstract} 
The article presents an experimental study of an issue of whether the formation of arrays of Ge quantum dots on the Si(001) surface is an equilibrium process or it is kinetically controlled. We deposited Ge on Si(001) at the room temperature and explored crystallization of the disordered Ge film as a result of annealing at 600{\,\textcelsius}.
The experiment has demonstrated that  the Ge/Si(001) film formed in the conditions of an isolated system consists of the standard patched wetting layer and large droplike clusters of Ge rather than of huts or domes which appear when a film is grown in a flux of Ge atoms arriving on its surface.
We conclude that the growth of the pyramids appearing at temperatures greater than 600\,{\textcelsius} is controlled by kinetics rather than thermodynamic equilibrium whereas the wetting layer is an equilibrium structure.
\end{abstract}




\end{abstractbox}
%

\end{frontmatter}




\section*{Introduction}
The issue of whether formation of arrays of Ge quantum dots on the Si(001) surface is an equilibrium driven or kinetically controlled process has not been solved since the very discovery of coherent GeSi and Ge islands by Eaglesham and Cerullo \cite{Eaglesham_Cerullo} and Mo {\it et al.} \cite{Mo}. 
 Numerous articles supported the model of kinetically-driven growth of Ge clusters  whereas others gave theoretical and experimental evidences of the equilibrium nature of Ge quantum dot arrays (see, e.\,g., Refs.\,\cite{Kinetically_Suppressed_Ripening, Kastner, Island_growth, Voigt_Review, Tersoff_Tromp, LLL, Goldfarb_2006, Goldfarb_2005, Facet-105, Stability-Anealing}).
In 1999, a detailed analysis of this long-standing (even at that time) issue was made by Shchukin and Bimberd in their now already classical review cited in Ref.\,\cite{Schukin-Bimberg} which covered a wide selection of various heteroepitaxial systems including Ge/Si(001) epitaxial films with Ge clusters.
Additional discussions of this issue can be found in a brief review section included to Ref.\,\cite{classification} or in our recent review article appeared in Ref.\,\cite{Yur_JNO}.

This article presents an experimental study of this issue. We deposited Ge on Si(001) at the room temperature and explored crystallization of the disordered Ge film as a result of a thermal treatment.
This experiment, originally conceived  as purely technological, gave somewhat unexpected results. First of all, it demonstrated that the Ge/Si(001) film formed in the conditions of a closed system consists of the usual patched wetting layer (WL) and large droplike clusters of Ge rather than huts or domes which appear when a film is deposited  in a flux of Ge atoms arriving on its surface. Additionally we observed nucleation of the $2\times 1$ reconstructed  phase (i.e. the ordered one) in the disordered Ge film deposited at the room temperature. And finally, we detected a mixture of $c(4\times 2)$ and  $p(2\times 2)$ reconstructions on the surface of the resultant Ge film annealed at high temperature (600\,\textcelsius) whereas the simultaneous presence of both these structures in comparable proportions on wetting layer patches is a distinctive feature of the low-temperature mode of the wetting layer growth ($<600$\,\textcelsius) \cite{initial_phase}.

Now, we proceed to the detailed presentation of the obtained results.$^{\rm a}$

\section*{Equipment, methods and samples}

Experiments were carried out using a specially built setup consisting of a ultrahigh-vacuum (UHV) molecular-beam epitaxy (MBE) chamber (Riber EVA~32) equipped with a RH20 reflected high-energy electron diffraction (RHED) unit (Staib Instruments) and connected with a UHV STM chamber (GPI~300) \cite{CMOS-compatible-EMRS,VCIAN2011,STM_GPI-Proc}. 

The Ge deposition rate and the Ge coverage ($h_{\rm Ge}$) were measured by a graduated in advance film thickness monitor (Inficon Leybold-Heraeus XTC 751-001-G1) with a quartz sensor installed at the MBE chamber. 
Samples were heated from the rear side by radiators of tantalum. 
The temperature was monitored with a tungsten-rhenium thermocouple mounted in the vacuum near the rear side of the samples and {\it in situ} graduated beforehand against  a specialized pyrometer (IMPAC IS 12-Si) which measured the Si sample temperature \textit{T} through a chamber window with an accuracy of \textpm $(0.003\,T$\,[{\textcelsius}] + 1)\,\textcelsius.
The atmosphere composition in the MBE chamber was monitored using a mass-spectrometer  residual gas analyzer (SRS RGA-200) before and during the process. Additional details concerning the used equipment can be found, e.\,g., in Refs.\,\cite{classification, CMOS-compatible-EMRS, VCIAN2011}.

Details of the pre-growth treatments of Si wafers, which included wet chemical etching and oxide removal by short high-temperature annealing ($T\sim$~900\,\textcelsius), can be found in our previous articles \cite{our_Si(001)_en,  stm-rheed-EMRS, phase_transition}. 
7\,\r{A} thick Ge films ($h_{\rm Ge}= 7$\,{\AA}) were deposited at the rate of 0.15\,{\AA}/s directly on the clean Si(001) surface at the room temperature. Then the samples were heated to 600{\,\textcelsius} at the maximum rate achievable for the used infrared heaters (0.24{\,\textcelsius}/s),$^{\rm b}$ 
annealed at this temperature for 5 minutes and cooled to the room temperature at the quenching mode \cite{hydrogenation_YUR} at the rate of 0.4{\,\textcelsius}/s. Afterwards the samples were moved into the STM chamber for the structural analysis. 
STM images were processed using the WSxM software \cite{WSxM}.
RHEED patterns were monitored on a screen and recorded with a video camera during the whole cycle of Ge film deposition and heat treatment.

\section*{Experimental data and their discussion}

\subsection*{A structure of the initial Ge film}

When Ge is deposited on the Si(001) surface at the room temperature
the $2\times 1$ RHEED pattern of Si(001) evolves into the $1\times 1$ one with the gradually decaying \textonehalf-streaks as the thickness of the deposited Ge film ($h_{\rm Ge}$) increases. However, even at $h_{\rm Ge}= 7$\,{\AA}, the {\textonehalf}-reflexes are still observable in the diffraction pattern (Fig.\,\ref{fig:RHEED_before}\,a,\,b) that is a direct indication of presence of the $2\times 1$ reconstructed domains of the film surface which occupy a part of its area. At the same time, the visible diffuse scattering of electrons and widening of the streaks indicates that the film is significantly disordered. So, we should conclude that an ordered $2\times 1$ reconstructed crystalline phase, although occupying a minor part of the surface, forms in the disordered Ge film deposited on Si(001) at the room temperature.

STM images obtained from the as-deposited Ge film demonstrate a highly disordered granular structure (Fig.\,\ref{fig:RHEED_before}\,c). The film thickness reaches 1\,nm due to its porosity. We failed to resolve the $2\times 1$ reconstructed islands or areas. However this does not allow one to state that such reconstructed domains are absent since the atomic resolution necessary for their observation was not reached at the mainly disordered surface of the  granular (and porous) Ge film. The RHEED patterns themselves demonstrating such reconstruction are a sufficient evidence for its presence on the surface. 

\subsection*{A structure of the {Ge/Si(001)} film after annealing}

\subsubsection*{The wetting layer}

RHEED patterns of the Ge layer drastically change as a result of sample annealing (Fig.\,\ref{fig:RHEED_after}): streaks of the $2 \times 1$ structure as well as 3D features are now observed. The 3D pattern becomes visible during sample heating at the temperature some higher than 500{\,\textcelsius}; the pronounced $2 \times 1$ structure appears, as it has already been shown previously \cite{Yur_JNO,SPIE_QD-chains}, during cooling at the temperature less then 600{\,\textcelsius}. So, we can conclude that the disordered Ge film deposited at the room temperature on Si(001) transforms as a result of annealing at 600{\,\textcelsius} into the ordered  $2 \times 1$ reconstructed Ge layer.

Furthermore, STM images of the resultant surface  demonstrate  typical pictures of the $M \times N$ reconstruction (Fig.\,\ref{fig:WL}): the surface is composed by patches bounded by the dimer-row vacancies and dimer vacancy lines \cite{Wu,Iwawaki_initial,Voigt,Wetting,Vailionis}.
There is no visible difference between this wetting layer, obtained as a result of annealing of the disordered Ge film, and wetting layers grown by molecular beam epitaxy at different conditions, both in low-temperature and high-temperature modes (Fig.\,\ref{fig:WL-360_600_650}); the only specific feature of this wetting layer is that, being formed at high temperature, it contains both $p(2\times 2)$ and $c(4\times 2)$ structures whereas only (or nearly only) the latter one is observed on the wetting layers deposited at high temperatures and a mixture of the two reconstructions is a characteristic feature of the wetting layers deposited at low temperatures \cite{Yur_JNO,SPIE_QD-chains, Growing_Ge_hut-structure,Nucleation_high-temperatures}.
But in general this minor peculiarity does not seem to be very significant in this case. 

Thus, we can conclude from the above that formation of the Ge/Si(001) wetting layer is controlled by system evolution to thermodynamic equilibrium, i.e. it is an equilibrium system.

\subsubsection*{Ge hillocks}

The thickness of the Ge/Si(001) wetting layer is known to be equal to 4 monolayers that is a little greater than 5.5\,\AA. We deposited 7\,\r{A} of Ge. Usually, if Ge is deposited at low temperatures (say, e.g., at {\em T}$_{\rm gr}$ = 360\,\textcelsius), at such coverages the excess Ge atoms form the well recognizable structure of the hut array (Fig.\,\ref{fig:array-600_360C}\,a) \cite{Mo,CMOS-compatible-EMRS,Yur_JNO}; if Ge is deposited at 600\,{\textcelsius} (or at the high-temperature mode) the excess Ge starts to gather into cluster arrays at even less coverages and forms pyramids in which split edges are often observed (Fig.\,\ref{fig:array-600_360C}\,b) \cite{VCIAN-2012, Prepyramid-to-pyramid, Morphological_evolution, Nucleation_high-temperatures}.
The question is where the excess Ge deposited at the room temperature flows during the heat treatment at 600\,{\textcelsius}. As we have observed, at these treatment Ge forms neither hut arrays nor pyramids or domes but it is gathered into large partially faceted droplike  clusters (Fig.\,\ref{fig:clusters}). The lateral dimensions of these oval hillocks vary from about 100 to about 200\,nm, their heights vary from nearly 10 to more than 20\,nm and their number density makes (1.2 to 1.3)$\times 10^9$\,cm$^{-2}$. ``Pelerines'' of multiple incomplete facets seen around apexes (Fig.\,\ref{fig:clusters}\,b) and indicating that these clusters grow from tops to bottoms \cite{VCIAN-2012}, 
as well as split edges (Fig.\,\ref{fig:clusters}\,b,\,d), demonstrate that they have some common features with the pyramids growing at the same temperature during the molecular beam epitaxy (Fig.\,\ref{fig:array-600_360C}\,b).

The presented observation allows us to definitely state that no huts (or domes) appear if a Ge/Si(001) film forms moving to equilibrium as a closed system (or, in fact, as an isolated system since outflow and radiative inflow of energy are balanced). Pyramid arrays (no wedges arise at high temperatures \cite{Nucleation_high-temperatures}) emerge only as a result of a process requiring Ge atoms to flow on the surface. 
So, this experiment unambiguously demonstrates that at least the growth of the high-temperature pyramids appearing at temperatures greater than 600\,{\textcelsius} is controlled by kinetics rather than evolution to thermodynamic equilibrium; equilibrium clusters are the oval ones.

Reasoning which allows us to come to this conclusion is very simple. 
The question is about the driving force of Ge cluster formation: if it is kinetics (random migration of atoms and their interactions) or thermodynamics (evolution of a system to equilibrium). The Ge/Si(001) isolated system comes to a thermodynamic equilibrium state for the given temperature as a result of isothermal annealing \cite{Ehrenfest-Afanassjewa_Thermodynamics}. If the pyramids would be equilibrium (or thermodynamically driven) structures they would appear as a result of annealing; or in other words, if they do not appear they are not equilibrium at high-temperature growth. We can conclude now that at high temperatures pyramids arise due to kinetics rather than thermodynamics, and thermodynamics would  give rise to the oval clusters instead of the pyramids.

It should be noted that we consider the oval mounds as equilibrium structures (or maybe close to equilibrium, since this does not matter for our conclusions)  despite that the annealing time seems too short. We believe that this conclusion is correct because the process, which first would  result in  appearance of large oval mounds on WL and then in their dissolution or reformation into some other structures, seems to be very unlikely in a closed system. Also we  can hardly assume that the oval cluster formation is constrained by some other kinetic processes, different from those constraining the pyramid formation during MBE, because the closed system that we consider (which is the isolated system at the phase of isothermal annealing) should evolve to its thermodynamic equilibrium and even if the oval mounds are not yet completely equilibrium formations they are certainly on this way and they likely may become larger or differently faceted in equilibrium but they may not become pyramids similar to those appearing as a result of MBE.

Now let us consider in a few words  the pyramid formation in the process of MBE. There are two competing kinetic processes on WL during MBE at the temperature of the described experiment: a process of a pyramid formation near a place of Ge atom ``landing'' on WL requiring a small migration length and a process of a long-distance migration of the ``landed'' Ge atoms along the WL surface resulting in appearance of oval clusters (whose number density is only about 10$^9$\,cm$^{-2}$). Probably, the latter process would be favorable at very low Ge deposition rates, approaching zero, when the system resembles the closed one and local supersaturation does not appear or has enough time to disappear due to adatom migration. At practical deposition rates, local supersaturation rises rapidly, Ge adatoms are incorporated by WL;  WL patches, whose upper layers are unstrained \cite{Yur_JNO,SPIE_QD-chains,Growing_Ge_hut-structure}, can no longer consume Ge and the strain would begin to grow unless Ge atoms start to form pyramids on patches as described by our model presented in Ref.\,\cite{Growing_Ge_hut-structure}. So, kinetics governs the processes of nucleation and growth of the high-temperature pyramids; otherwise the excess Ge adatoms would be distributed into the observed oval clusters as it is prescribed by thermodynamics.

We should note also that unlike {SiGe} islands grown by {Ge} diffusion from a local source \cite{Annealing_SiGe_600C_Montalenti,Annealing_SiGe_600C_nanotechnology} at the temperatures close or equal to the temperature of the discussed experiment, which also have oval shapes although some different from those described in this article, the oval mounds described by us evolve to equilibrium with the wetting layer. The islands studied in Refs.\,\cite{Annealing_SiGe_600C_Montalenti,Annealing_SiGe_600C_nanotechnology} arise and grow in a permanent flux of adatoms directionally migrating along the wetting layer from the Ge stripe (the source of Ge) to the Si surface free of Ge. Consequently, they never come to equilibrium with the wetting layer (although they grow under the isothermal annealing) and their formation is controlled by kinetic processes, i.\,e. by diffusion and capture of diffusing atoms.

\section*{Conclusion}

In conclusion of the article, we would like to emphasize its main statements.

First, we conclude that an ordered $2\times 1$ reconstructed crystalline phase forms in the disordered Ge film during its deposition on the Si(001) surface at the room temperature.
Second, as a result of annealing at 600{\,\textcelsius} the disordered Ge film deposited at the room temperature on Si(001) transforms  into the ordered  $2 \times 1$ reconstructed Ge wetting layer. 
Third, the resultant wetting layer surface is $M \times N$ reconstructed, i.e. it consists of patches bounded by the dimer-row vacancies and dimer vacancy lines; apexes of the patches are formed by both $p(2\times 2)$ and $c(4\times 2)$ structures.
Fourth, huts (pyramids or wedges) or domes do not emerge if a Ge/Si(001) film is formed in a closed system by annealing at 600{\,\textcelsius} of a Ge layer deposited on Si(001) at the room temperature; the excess Ge atoms from the initial Ge film are gathered into large partially faceted oval clusters.

And finally, we have demonstrated that at least the growth of the pyramids appearing at temperatures greater than 600\,{\textcelsius} is controlled by kinetics rather than evolution to thermodynamic equilibrium whereas the wetting layer is an equilibrium structure. Consequently, we can conclude that these nonequilibrium, kinetically controlled, Ge quantum dots grow on the equilibrium, thermodynamically controlled, Ge/Si(001) wetting layer. Thus, the wetting layer could determine only the places of their nucleation and maybe the structure of their nuclei but cannot control their growth process; this inference completely agrees with the hut nucleation and growth scenarios proposed in our previous publications \cite{Yur_JNO, SPIE_QD-chains, Growing_Ge_hut-structure}.



\begin{backmatter}

\section*{Competing interests}
  The authors declare that they have no competing interests.

\section*{Author's contributions}
MSS conceived of the study and designed it, he carried out the experiments, and took part in discussions and interpretation of the results. 
LVA participated in the design of the study, and carried out the experiments; she took part in discussions and interpretation of the results. 
VAY performed data analysis and interpreted the results, he drafted the manuscript. 
All authors read and approved
the final manuscript.

\section*{Acknowledgements}
  We thank Ms. N. V. Kiryanova for her invaluable contribution to arrangement of this research.
We also thank Ms. L. M. Krylova for chemical treatments of the samples.
Equipment of the Center for Collective Use of Scientific Equipment of GPI RAS was used for this study. We acknowledge the support of this research.

\section*{Abbreviations}
MBE: molecular beam epitaxy;
RHEED: reflected high-energy electron diffraction;
WL: wetting layer

\section*{Endnotes}
$^{\rm a}$A preprint of the current article is cited in Ref.\,\cite{Anneal@600C_arXiv}.

$^{\rm b}$The heating process does not seem to be really adiabatic: its rate was insufficiently high and we observed changes in the surface structure that means that some changes in entropy took place. However, we performed heating as rapidly as we could to minimize its effect on the Ge layer.


\bibliographystyle{bmc-mathphys} 
\bibliography{Literature_on_Si-Ge}      


\begin{thebibliography}{39}
\ifx \bisbn   \undefined \def \bisbn  #1{ISBN #1}\fi
\ifx \binits  \undefined \def \binits#1{#1}\fi
\ifx \bauthor  \undefined \def \bauthor#1{#1}\fi
\ifx \batitle  \undefined \def \batitle#1{#1}\fi
\ifx \bjtitle  \undefined \def \bjtitle#1{#1}\fi
\ifx \bvolume  \undefined \def \bvolume#1{\textbf{#1}}\fi
\ifx \byear  \undefined \def \byear#1{#1}\fi
\ifx \bissue  \undefined \def \bissue#1{#1}\fi
\ifx \bfpage  \undefined \def \bfpage#1{#1}\fi
\ifx \blpage  \undefined \def \blpage #1{#1}\fi
\ifx \burl  \undefined \def \burl#1{\textsf{#1}}\fi
\ifx \doiurl  \undefined \def \doiurl#1{\textsf{#1}}\fi
\ifx \betal  \undefined \def \betal{\textit{et al.}}\fi
\ifx \binstitute  \undefined \def \binstitute#1{#1}\fi
\ifx \binstitutionaled  \undefined \def \binstitutionaled#1{#1}\fi
\ifx \bctitle  \undefined \def \bctitle#1{#1}\fi
\ifx \beditor  \undefined \def \beditor#1{#1}\fi
\ifx \bpublisher  \undefined \def \bpublisher#1{#1}\fi
\ifx \bbtitle  \undefined \def \bbtitle#1{#1}\fi
\ifx \bedition  \undefined \def \bedition#1{#1}\fi
\ifx \bseriesno  \undefined \def \bseriesno#1{#1}\fi
\ifx \blocation  \undefined \def \blocation#1{#1}\fi
\ifx \bsertitle  \undefined \def \bsertitle#1{#1}\fi
\ifx \bsnm \undefined \def \bsnm#1{#1}\fi
\ifx \bsuffix \undefined \def \bsuffix#1{#1}\fi
\ifx \bparticle \undefined \def \bparticle#1{#1}\fi
\ifx \barticle \undefined \def \barticle#1{#1}\fi
\ifx \bconfdate \undefined \def \bconfdate #1{#1}\fi
\ifx \botherref \undefined \def \botherref #1{#1}\fi
\ifx \url \undefined \def \url#1{\textsf{#1}}\fi
\ifx \bchapter \undefined \def \bchapter#1{#1}\fi
\ifx \bbook \undefined \def \bbook#1{#1}\fi
\ifx \bcomment \undefined \def \bcomment#1{#1}\fi
\ifx \oauthor \undefined \def \oauthor#1{#1}\fi
\ifx \citeauthoryear \undefined \def \citeauthoryear#1{#1}\fi
\ifx \endbibitem  \undefined \def \endbibitem {}\fi
\ifx \bconflocation  \undefined \def \bconflocation#1{#1}\fi
\ifx \arxivurl  \undefined \def \arxivurl#1{\textsf{#1}}\fi
\csname PreBibitemsHook\endcsname

\bibitem{Eaglesham_Cerullo}
\begin{barticle}
\bauthor{\bsnm{Eaglesham}, \binits{D.J.}},
\bauthor{\bsnm{Cerullo}, \binits{M.}}:
\batitle{{Dislocation-free Stranski-Krastanow growth of Ge on Si(100)}}.
\bjtitle{Phys. Rev. Lett.}
\bvolume{64}(\bissue{16}),
\bfpage{1943}
(\byear{1990})
\end{barticle}
\endbibitem

\bibitem{Mo}
\begin{barticle}
\bauthor{\bsnm{Mo}, \binits{Y.-W.}},
\bauthor{\bsnm{Savage}, \binits{D.E.}},
\bauthor{\bsnm{Swartzentruber}, \binits{B.S.}},
\bauthor{\bsnm{Lagally}, \binits{M.G.}}:
\batitle{{Kinetic pathway in Stranski-Krastanov growth of Ge on Si(001)}}.
\bjtitle{Phys. Rev. Lett.}
\bvolume{65},
\bfpage{1020}
(\byear{1990})
\end{barticle}
\endbibitem

\bibitem{Kinetically_Suppressed_Ripening}
\begin{barticle}
\bauthor{\bsnm{McKay}, \binits{M.R.}},
\bauthor{\bsnm{Venables}, \binits{J.A.}},
\bauthor{\bsnm{Drucker}, \binits{J.}}:
\batitle{{Kinetically suppressed Ostwald ripening of Ge/Si(100) hut clusters}}.
\bjtitle{Phys. Rev. Lett.}
\bvolume{101},
\bfpage{216104}
(\byear{2008})
\end{barticle}
\endbibitem

\bibitem{Kastner}
\begin{barticle}
\bauthor{\bsnm{K{\"{a}}stner}, \binits{M.}},
\bauthor{\bsnm{Voigtl{\"{a}}nder}, \binits{B.}}:
\batitle{{Kinetically self-limiting growth of Ge islands on Si(001)}}.
\bjtitle{Phys. Rev. Lett.}
\bvolume{82}(\bissue{13}),
\bfpage{2745}
(\byear{1999})
\end{barticle}
\endbibitem

\bibitem{Island_growth}
\begin{barticle}
\bauthor{\bsnm{Jesson}, \binits{D.E.}},
\bauthor{\bsnm{Chen}, \binits{G.}},
\bauthor{\bsnm{Chen}, \binits{K.M.}},
\bauthor{\bsnm{Pennycook}, \binits{S.J.}}:
\batitle{{Self-limiting growth of strained faceted islands}}.
\bjtitle{Phys. Rev. Lett.}
\bvolume{80}(\bissue{23}),
\bfpage{5156}
(\byear{1998})
\end{barticle}
\endbibitem

\bibitem{Voigt_Review}
\begin{barticle}
\bauthor{\bsnm{Voigtl{\"{a}}nder}, \binits{B.}}:
\batitle{{Fundamental processes in Si/Si and Ge/Si epitaxy studied by scanning
  tunneling microscopy during growth}}.
\bjtitle{Surf. Sci. Rep.}
\bvolume{43},
\bfpage{127}
(\byear{2001})
\end{barticle}
\endbibitem

\bibitem{Tersoff_Tromp}
\begin{barticle}
\bauthor{\bsnm{Tersoff}, \binits{J.}},
\bauthor{\bsnm{Tromp}, \binits{R.M.}}:
\batitle{{Shape transitions in growth of strained islands: Spontaneous
  formation of quantum wires}}.
\bjtitle{Phys. Rev. Lett.}
\bvolume{70}(\bissue{18}),
\bfpage{2782}
(\byear{1993})
\end{barticle}
\endbibitem

\bibitem{LLL}
\begin{barticle}
\bauthor{\bsnm{Li}, \binits{A.}},
\bauthor{\bsnm{Liu}, \binits{F.}},
\bauthor{\bsnm{Lagally}, \binits{M.G.}}:
\batitle{{Equilibrium shape of two-dimensional islands under stress}}.
\bjtitle{Phys. Rev. Lett.}
\bvolume{85}(\bissue{9}),
\bfpage{1922}
(\byear{2000})
\end{barticle}
\endbibitem

\bibitem{Goldfarb_2006}
\begin{barticle}
\bauthor{\bsnm{Goldfarb}, \binits{I.}},
\bauthor{\bsnm{Banks-Sills}, \binits{L.}},
\bauthor{\bsnm{Eliasi}, \binits{R.}}:
\batitle{{Is the elongation of Ge huts in the low-temperature regime governed
  by kinetics?}}
\bjtitle{Phys. Rev. Lett.}
\bvolume{97},
\bfpage{206101}
(\byear{2006})
\end{barticle}
\endbibitem

\bibitem{Goldfarb_2005}
\begin{barticle}
\bauthor{\bsnm{Goldfarb}, \binits{I.}}:
\batitle{{Effect of strain on the appearance of subcritical nuclei of Ge
  nanohuts on Si(001)}}.
\bjtitle{Phys. Rev. Lett.}
\bvolume{95},
\bfpage{025501}
(\byear{2005})
\end{barticle}
\endbibitem

\bibitem{Facet-105}
\begin{barticle}
\bauthor{\bsnm{Raiteri}, \binits{P.}},
\bauthor{\bsnm{Migas}, \binits{D.B.}},
\bauthor{\bsnm{Miglio}, \binits{L.}},
\bauthor{\bsnm{Rastelli}, \binits{A.}},
\bauthor{\bparticle{von} \bsnm{K{\"{a}}nel}, \binits{H.}}:
\batitle{{Critical role of the surface reconstruction in the thermodynamic
  stability of $\{105\}$ Ge pyramids on Si(001)}}.
\bjtitle{Phys. Rev. Lett.}
\bvolume{88}(\bissue{25}),
\bfpage{256103}
(\byear{2002})
\end{barticle}
\endbibitem

\bibitem{Stability-Anealing}
\begin{barticle}
\bauthor{\bsnm{Medeiros-Ribeiro}, \binits{G.}},
\bauthor{\bsnm{Kamins}, \binits{T.I.}},
\bauthor{\bsnm{Ohlberg}, \binits{D.A.A.}},
\bauthor{\bsnm{Williams}, \binits{R.S.}}:
\batitle{{Annealing of Ge nanocrystals on Si(001) at 550$^{\circ}$C:
  Metastability of huts and the stability of pyramids and domes}}.
\bjtitle{Phys. Rev. B}
\bvolume{58}(\bissue{7}),
\bfpage{3533}
(\byear{1998})
\end{barticle}
\endbibitem

\bibitem{Schukin-Bimberg}
\begin{barticle}
\bauthor{\bsnm{Schukin}, \binits{V.A.}},
\bauthor{\bsnm{Bimberg}, \binits{D.}}:
\batitle{{Spontaneous ordering of nanostructures on crystal surfaces}}.
\bjtitle{Rev. Modern Phys.}
\bvolume{71}(\bissue{4}),
\bfpage{1125}
(\byear{1999})
\end{barticle}
\endbibitem

\bibitem{classification}
\begin{barticle}
\bauthor{\bsnm{Arapkina}, \binits{L.V.}},
\bauthor{\bsnm{Yuryev}, \binits{V.A.}}:
\batitle{{Classification of Ge hut clusters in arrays formed by molecular beam
  epitaxy at low temperatures on the Si(001) surface}}.
\bjtitle{Phys. Usp.}
\bvolume{53}(\bissue{3}),
\bfpage{279}
(\byear{2010}).
\end{barticle}
\endbibitem

\bibitem{Yur_JNO}
\begin{barticle}
\bauthor{\bsnm{Yuryev}, \binits{V.A.}},
\bauthor{\bsnm{Arapkina}, \binits{L.V.}},
\bauthor{\bsnm{Storozhevykh}, \binits{M.S.}},
\bauthor{\bsnm{Uvarov}, \binits{O.V.}},
\bauthor{\bsnm{Kalinushkin}, \binits{V.P.}}:
\batitle{{Ge/Si heterostructures with dense chains of stacked quantum dots}}.
\bjtitle{J. Nanoelectron. Optoelectron.}
\bvolume{9}(\bissue{2}),
\bfpage{196}--\blpage{218}
(\byear{2014})
\end{barticle}
\endbibitem

\bibitem{initial_phase}
\begin{barticle}
\bauthor{\bsnm{Arapkina}, \binits{L.V.}},
\bauthor{\bsnm{Yuryev}, \binits{V.A.}}:
\batitle{{An initial phase of Ge hut array formation at low temperature on
  Si(001)}}.
\bjtitle{J. Appl. Phys.}
\bvolume{109}(\bissue{10}),
\bfpage{104319}
(\byear{2011}).
\end{barticle}
\endbibitem

\bibitem{CMOS-compatible-EMRS}
\begin{barticle}
\bauthor{\bsnm{Arapkina}, \binits{L.V.}},
\bauthor{\bsnm{Yuryev}, \binits{V.A.}}:
\batitle{{CMOS compatible dense arrays of Ge quantum dots on the Si(001)
  surface: Hut cluster nucleation, atomic structure, and array life cycle
  during UHV MBE growth}}.
\bjtitle{Nanoscale Res. Lett.}
\bvolume{6},
\bfpage{345}
(\byear{2011}).
\end{barticle}
\endbibitem

\bibitem{VCIAN2011}
\begin{barticle}
\bauthor{\bsnm{Yuryev}, \binits{V.A.}},
\bauthor{\bsnm{Arapkina}, \binits{L.V.}}:
\batitle{{Ge quantum dot arrays grown by ultrahigh vacuum molecular-beam
  epitaxy on the Si(001) surface: nucleation, morphology, and CMOS
  compatibility}}.
\bjtitle{Nanoscale Res. Lett.}
\bvolume{6},
\bfpage{522}
(\byear{2011}).
\end{barticle}
\endbibitem

\bibitem{STM_GPI-Proc}
\begin{bchapter}
\bauthor{\bsnm{Eltsov}, \binits{K.N.}},
\bauthor{\bsnm{Klimov}, \binits{A.N.}},
\bauthor{\bsnm{Kosyakov}, \binits{A.N.}},
\bauthor{\bsnm{Obyedkov}, \binits{O.V.}},
\bauthor{\bsnm{Shevlyuga}, \binits{V.M.}},
\bauthor{\bsnm{Yurov}, \binits{V.Y.}}:
\bctitle{{Ultrahigh vacuum scanning funnelling microscope GPI-300}}.
In: \beditor{\bsnm{Konov}, \binits{V.I.}},
\beditor{\bsnm{Eltsov}, \binits{K.N.}} (eds.)
\bbtitle{Chemical State and Atomic Structure of Fcc Metal Surfaces in Chemical
  Reaction with Halogens}.
\bsertitle{Proc. of General Physics Institute},
vol. \bseriesno{59},
p. \bfpage{45}.
\bpublisher{Nauka},
\blocation{Moscow, Russia}
(\byear{2003})
\end{bchapter}
\endbibitem

\bibitem{our_Si(001)_en}
\begin{barticle}
\bauthor{\bsnm{Arapkina}, \binits{L.V.}},
\bauthor{\bsnm{Shevlyuga}, \binits{V.M.}},
\bauthor{\bsnm{Yuryev}, \binits{V.A.}}:
\batitle{{Structure and peculiarities of the $(8\times n)$-type Si(001) surface
  prepared in a molecular beam epitaxy chamber: A scanning tunneling microscopy
  study}}.
\bjtitle{JETP Lett.}
\bvolume{87},
\bfpage{215}
(\byear{2008}).
\end{barticle}
\endbibitem

\bibitem{stm-rheed-EMRS}
\begin{barticle}
\bauthor{\bsnm{Arapkina}, \binits{L.V.}},
\bauthor{\bsnm{Yuryev}, \binits{V.A.}},
\bauthor{\bsnm{Chizh}, \binits{K.V.}},
\bauthor{\bsnm{Shevlyuga}, \binits{V.M.}},
\bauthor{\bsnm{Storojevyh}, \binits{M.S.}},
\bauthor{\bsnm{Krylova}, \binits{L.A.}}:
\batitle{{Phase transition on the Si(001) clean surface prepared in UHV MBE
  chamber: A study by high resolution STM and in situ RHEED}}.
\bjtitle{Nanoscale Res. Lett.}
\bvolume{6},
\bfpage{218}
(\byear{2011}).
\end{barticle}
\endbibitem

\bibitem{phase_transition}
\begin{barticle}
\bauthor{\bsnm{Arapkina}, \binits{L.V.}},
\bauthor{\bsnm{Yuryev}, \binits{V.A.}},
\bauthor{\bsnm{Shevlyuga}, \binits{V.M.}},
\bauthor{\bsnm{Chizh}, \binits{K.V.}}:
\batitle{{Phase transition between ${(2\times 1)}$ and ${c(8\times 8)}$
  reconstructions observed on the Si(001) surface around 600$^{\circ}$C}}.
\bjtitle{JETP Lett.}
\bvolume{92}(\bissue{5}),
\bfpage{310}
(\byear{2010}).
\end{barticle}
\endbibitem

\bibitem{hydrogenation_YUR}
\begin{barticle}
\bauthor{\bsnm{Arapkina}, \binits{L.V.}},
\bauthor{\bsnm{Krylova}, \binits{L.A.}},
\bauthor{\bsnm{Chizh}, \binits{K.V.}},
\bauthor{\bsnm{Chapnin}, \binits{V.A.}},
\bauthor{\bsnm{Uvarov}, \binits{O.V.}},
\bauthor{\bsnm{Yuryev}, \binits{V.A.}}:
\batitle{{Application of hydrogenation to low-temperature cleaning of the
  Si(001) surface in the processes of molecular-beam epitaxy: Investigation by
  scanning tunneling microscopy, reflected high-energy electron diffraction,
  and high resolution transmission electron microscopy}}.
\bjtitle{J. Appl. Phys.}
\bvolume{112}(\bissue{1}),
\bfpage{014311}
(\byear{2012}).
\end{barticle}
\endbibitem

\bibitem{WSxM}
\begin{barticle}
\bauthor{\bsnm{Horcas}, \binits{I.}},
\bauthor{\bsnm{Fernandez}, \binits{R.}},
\bauthor{\bsnm{Gomez-Rodriguez}, \binits{J.M.}},
\bauthor{\bsnm{Colchero}, \binits{J.}},
\bauthor{\bsnm{Gomez-Herrero}, \binits{J.}},
\bauthor{\bsnm{Baro}, \binits{A.M.}}:
\batitle{{WSxM: A software for scanning probe microscopy and a tool for
  nanotechnology}}.
\bjtitle{Rev. Sci. Instrum.}
\bvolume{78},
\bfpage{013705}
(\byear{2007})
\end{barticle}
\endbibitem

\bibitem{SPIE_QD-chains}
\begin{bchapter}
\bauthor{\bsnm{Yuryev}, \binits{V.A.}},
\bauthor{\bsnm{Arapkina}, \binits{L.V.}},
\bauthor{\bsnm{Storozhevykh}, \binits{M.S.}},
\bauthor{\bsnm{Uvarov}, \binits{O.V.}},
\bauthor{\bsnm{Kalinushkin}, \binits{V.P.}}:
\bctitle{{Dense chains of stacked quantum dots in Ge/Si heterostructures}}.
In: \beditor{\bsnm{Adelung}, \binits{R.}} (ed.)
\bbtitle{Nanotechnology VI}.
\bsertitle{Proc. SPIE},
vol. \bseriesno{8766},
p. \bfpage{876606}.
\bconflocation{SPIE, Bellingham, WA}
(\byear{2013}).
\bcomment{{SPIE Paper Number 8766-5}}
\end{bchapter}
\endbibitem

\bibitem{Wu}
\begin{barticle}
\bauthor{\bsnm{Wu}, \binits{F.}},
\bauthor{\bsnm{Chen}, \binits{X.}},
\bauthor{\bsnm{Zhang}, \binits{Z.}},
\bauthor{\bsnm{Lagally}, \binits{M.G.}}:
\batitle{{Reversal of step roughness on Ge-covered vicinal Si(001)}}.
\bjtitle{Phys. Rev. Lett.}
\bvolume{74},
\bfpage{574}
(\byear{1995})
\end{barticle}
\endbibitem

\bibitem{Iwawaki_initial}
\begin{barticle}
\bauthor{\bsnm{Iwawaki}, \binits{F.}},
\bauthor{\bsnm{Tomitori}, \binits{M.}},
\bauthor{\bsnm{Nishikawa}, \binits{O.}}:
\batitle{{STM study of initial stage of Ge epitaxy on Si(001)}}.
\bjtitle{Ultramicroscopy}
\bvolume{42--44},
\bfpage{902}
(\byear{1992})
\end{barticle}
\endbibitem

\bibitem{Voigt}
\begin{barticle}
\bauthor{\bsnm{Voigtl{\"{a}}nder}, \binits{B.}},
\bauthor{\bsnm{K{\"{a}}stner}, \binits{M.}}:
\batitle{{Evolution of the strain relaxation in a Ge layer on Si(001) by
  reconstruction and intermixing}}.
\bjtitle{Phys. Rev. B}
\bvolume{60},
\bfpage{5121}
(\byear{1999})
\end{barticle}
\endbibitem

\bibitem{Wetting}
\begin{barticle}
\bauthor{\bsnm{Migas}, \binits{D.B.}},
\bauthor{\bsnm{Raiteri}, \binits{P.}},
\bauthor{\bsnm{Miglio}, \binits{L.}},
\bauthor{\bsnm{Rastelli}, \binits{A.}},
\bauthor{\bparticle{von} \bsnm{K{\"{a}}nel}, \binits{H.}}:
\batitle{{Evolution of Ge/Si(001) wetting layer during Si overgrowth and
  crossover between thermodynamic and kinetic behavior}}.
\bjtitle{Phys. Rev. B}
\bvolume{69},
\bfpage{235318}
(\byear{2004})
\end{barticle}
\endbibitem

\bibitem{Vailionis}
\begin{barticle}
\bauthor{\bsnm{Vailionis}, \binits{A.}},
\bauthor{\bsnm{Cho}, \binits{B.}},
\bauthor{\bsnm{Glass}, \binits{G.}},
\bauthor{\bsnm{Desjardins}, \binits{P.}},
\bauthor{\bsnm{Cahill}, \binits{D.G.}},
\bauthor{\bsnm{Greene}, \binits{J.E.}}:
\batitle{{Pathway to the strain-driven two-dimensional to three-dimensional
  transitions during growth of Ge on Si(001)}}.
\bjtitle{Phys. Rev. Lett.}
\bvolume{85},
\bfpage{3672}--\blpage{3675}
(\byear{2000})
\end{barticle}
\endbibitem

\bibitem{Growing_Ge_hut-structure}
\begin{barticle}
\bauthor{\bsnm{Arapkina}, \binits{L.V.}},
\bauthor{\bsnm{Yuryev}, \binits{V.A.}}:
\batitle{{On atomic structure of Ge huts growing on the Ge/Si(001) wetting
  layer}}.
\bjtitle{J. Appl. Phys.}
\bvolume{114},
\bfpage{104304}
(\byear{2013}).
\end{barticle}
\endbibitem

\bibitem{Nucleation_high-temperatures}
\begin{barticle}
\bauthor{\bsnm{Yuryev}, \binits{V.A.}},
\bauthor{\bsnm{Arapkina}, \binits{L.V.}}:
\batitle{{Nucleation of Ge clusters at high temperatures on Ge/Si(001) wetting
  layer}}.
\bjtitle{J. Appl. Phys.}
\bvolume{111},
\bfpage{094307}
(\byear{2012}).
\end{barticle}
\endbibitem

\bibitem{VCIAN-2012}
\begin{barticle}
\bauthor{\bsnm{Yuryev}, \binits{V.A.}},
\bauthor{\bsnm{Arapkina}, \binits{L.V.}},
\bauthor{\bsnm{Storozhevykh}, \binits{M.S.}},
\bauthor{\bsnm{Chapnin}, \binits{V.A.}},
\bauthor{\bsnm{Chizh}, \binits{K.V.}},
\bauthor{\bsnm{Uvarov}, \binits{O.V.}},
\bauthor{\bsnm{Kalinushkin}, \binits{V.P.}},
\bauthor{\bsnm{Zhukova}, \binits{E.S.}},
\bauthor{\bsnm{Prokhorov}, \binits{A.S.}},
\bauthor{\bsnm{Spektor}, \binits{I.E.}},
\bauthor{\bsnm{Gorshunov}, \binits{B.P.}}:
\batitle{{Ge/Si(001) heterostructures with dense arrays of Ge quantum dots:
  morphology, defects, photo-emf spectra and terahertz conductivity}}.
\bjtitle{Nanoscale Res. Lett.}
\bvolume{7},
\bfpage{414}
(\byear{2012}).
\end{barticle}
\endbibitem

\bibitem{Prepyramid-to-pyramid}
\begin{barticle}
\bauthor{\bsnm{Rastelli}, \binits{A.}},
\bauthor{\bparticle{von} \bsnm{K{\"{a}}nel}, \binits{H.}},
\bauthor{\bsnm{Spencer}, \binits{B.J.}},
\bauthor{\bsnm{Tersoff}, \binits{J.}}:
\batitle{{Prepyramid-to-pyramid transition of SiGe islands on Si(001)}}.
\bjtitle{Phys. Rev. B}
\bvolume{68},
\bfpage{115301}
(\byear{2003})
\end{barticle}
\endbibitem

\bibitem{Morphological_evolution}
\begin{barticle}
\bauthor{\bsnm{Grydlik}, \binits{M.}},
\bauthor{\bsnm{Brehm}, \binits{M.}},
\bauthor{\bsnm{Sch{\"{a}}ffler}, \binits{F.}}:
\batitle{{Morphological evolution of Ge/Si(001) quantum dot rings formed at the
  rim of wet-etched pits}}.
\bjtitle{Nanoscale Res. Lett.}
\bvolume{7},
\bfpage{601}
(\byear{2012})
\end{barticle}
\endbibitem

\bibitem{Ehrenfest-Afanassjewa_Thermodynamics}
\begin{bbook}
\bauthor{\bsnm{Ehrenfest-Afanassjewa}, \binits{T.}}:
\bbtitle{{Die Grundlagen der Thermodynamik}}.
\bpublisher{E.\,J. Brill},
\blocation{Leiden, Germany}
(\byear{1956})
\end{bbook}
\endbibitem

\bibitem{Annealing_SiGe_600C_Montalenti}
\begin{barticle}
\bauthor{\bsnm{Vanacore}, \binits{G.M.}},
\bauthor{\bsnm{Nicotra}, \binits{G.}},
\bauthor{\bsnm{Zani}, \binits{M.}},
\bauthor{\bsnm{Bollani}, \binits{M.}},
\bauthor{\bsnm{E.~Bonera}, \binits{F.M.}},
\bauthor{\bsnm{Capellini}, \binits{G.}},
\bauthor{\bsnm{Isella}, \binits{G.}},
\bauthor{\bsnm{Osmond}, \binits{J.}},
\bauthor{\bsnm{A.~Picco}, \binits{F.B.}},
\bauthor{\bsnm{Tagliaferri}, \binits{A.}}:
\batitle{{Delayed plastic relaxation limit in {SiGe} islands grown by {Ge}
  diffusion from a local source}}.
\bjtitle{J. Appl. Phys.}
\bvolume{117},
\bfpage{104309}
(\byear{2015})
\end{barticle}
\endbibitem

\bibitem{Annealing_SiGe_600C_nanotechnology}
\begin{barticle}
\bauthor{\bsnm{Vanacore}, \binits{G.M.}},
\bauthor{\bsnm{Zani}, \binits{M.}},
\bauthor{\bsnm{Bollani}, \binits{M.}},
\bauthor{\bsnm{Bonera}, \binits{E.}},
\bauthor{\bsnm{Nicotra}, \binits{G.}},
\bauthor{\bsnm{Osmond}, \binits{J.}},
\bauthor{\bsnm{Capellini}, \binits{G.}},
\bauthor{\bsnm{Isella}, \binits{G.}},
\bauthor{\bsnm{Tagliaferri}, \binits{A.}}:
\batitle{Monitoring the kinetic evolution of self-assembled {SiGe} islands
  grown by {Ge} surface thermal diffusion from a local source}.
\bjtitle{Nanotechnology}
\bvolume{25},
\bfpage{135606}
(\byear{2014})
\end{barticle}
\endbibitem

\bibitem{Anneal@600C_arXiv}
\begin{botherref}
\oauthor{\bsnm{Storozhevykh}, \binits{M.S.}},
\oauthor{\bsnm{Arapkina}, \binits{L.V.}},
\oauthor{\bsnm{Yuryev}, \binits{V.A.}}:
{Ge/Si(001) wetting layer formed by Ge deposition at room temperature followed
  by annealing at 600{\,\textcelsius} and its structural features}.
Posted on arXiv:1409.5422v2
(2014)
\end{botherref}
\endbibitem

\end{thebibliography}

\newcommand{\BMCxmlcomment}[1]{}

\BMCxmlcomment{

<refgrp>

<bibl id="B1">
  <title><p>{Dislocation-free Stranski-Krastanow growth of Ge on
  Si(100)}</p></title>
  <aug>
    <au><snm>Eaglesham</snm><fnm>D. J.</fnm></au>
    <au><snm>Cerullo</snm><fnm>M.</fnm></au>
  </aug>
  <source>Phys. Rev. Lett.</source>
  <pubdate>1990</pubdate>
  <volume>64</volume>
  <issue>16</issue>
  <fpage>1943</fpage>
</bibl>

<bibl id="B2">
  <title><p>{Kinetic pathway in Stranski-Krastanov growth of Ge on
  Si(001)}</p></title>
  <aug>
    <au><snm>Mo</snm><fnm>Y. W.</fnm></au>
    <au><snm>Savage</snm><fnm>D. E.</fnm></au>
    <au><snm>Swartzentruber</snm><fnm>B. S.</fnm></au>
    <au><snm>Lagally</snm><fnm>M. G.</fnm></au>
  </aug>
  <source>Phys. Rev. Lett.</source>
  <pubdate>1990</pubdate>
  <volume>65</volume>
  <fpage>1020</fpage>
</bibl>

<bibl id="B3">
  <title><p>{Kinetically suppressed Ostwald ripening of Ge/Si(100) hut
  clusters}</p></title>
  <aug>
    <au><snm>McKay</snm><fnm>M. R.</fnm></au>
    <au><snm>Venables</snm><fnm>J. A.</fnm></au>
    <au><snm>Drucker</snm><fnm>J.</fnm></au>
  </aug>
  <source>Phys. Rev. Lett.</source>
  <pubdate>2008</pubdate>
  <volume>101</volume>
  <fpage>216104</fpage>
</bibl>

<bibl id="B4">
  <title><p>{Kinetically self-limiting growth of Ge islands on
  Si(001)}</p></title>
  <aug>
    <au><snm>K{\"{a}}stner</snm><fnm>M.</fnm></au>
    <au><snm>Voigtl{\"{a}}nder</snm><fnm>B.</fnm></au>
  </aug>
  <source>Phys. Rev. Lett.</source>
  <pubdate>1999</pubdate>
  <volume>82</volume>
  <issue>13</issue>
  <fpage>2745</fpage>
</bibl>

<bibl id="B5">
  <title><p>{Self-limiting growth of strained faceted islands}</p></title>
  <aug>
    <au><snm>Jesson</snm><fnm>D. E.</fnm></au>
    <au><snm>Chen</snm><fnm>G.</fnm></au>
    <au><snm>Chen</snm><fnm>K. M.</fnm></au>
    <au><snm>Pennycook</snm><fnm>S. J.</fnm></au>
  </aug>
  <source>Phys. Rev. Lett.</source>
  <pubdate>1998</pubdate>
  <volume>80</volume>
  <issue>23</issue>
  <fpage>5156</fpage>
</bibl>

<bibl id="B6">
  <title><p>{Fundamental processes in Si/Si and Ge/Si epitaxy studied by
  scanning tunneling microscopy during growth}</p></title>
  <aug>
    <au><snm>Voigtl{\"{a}}nder</snm><fnm>B.</fnm></au>
  </aug>
  <source>Surf. Sci. Rep.</source>
  <pubdate>2001</pubdate>
  <volume>43</volume>
  <fpage>127</fpage>
</bibl>

<bibl id="B7">
  <title><p>{Shape transitions in growth of strained islands: Spontaneous
  formation of quantum wires}</p></title>
  <aug>
    <au><snm>Tersoff</snm><fnm>J.</fnm></au>
    <au><snm>Tromp</snm><fnm>R. M.</fnm></au>
  </aug>
  <source>Phys. Rev. Lett.</source>
  <pubdate>1993</pubdate>
  <volume>70</volume>
  <issue>18</issue>
  <fpage>2782</fpage>
</bibl>

<bibl id="B8">
  <title><p>{Equilibrium shape of two-dimensional islands under
  stress}</p></title>
  <aug>
    <au><snm>Li</snm><fnm>A.</fnm></au>
    <au><snm>Liu</snm><fnm>F.</fnm></au>
    <au><snm>Lagally</snm><fnm>M. G.</fnm></au>
  </aug>
  <source>Phys. Rev. Lett.</source>
  <pubdate>2000</pubdate>
  <volume>85</volume>
  <issue>9</issue>
  <fpage>1922</fpage>
</bibl>

<bibl id="B9">
  <title><p>{Is the elongation of Ge huts in the low-temperature regime
  governed by kinetics?}</p></title>
  <aug>
    <au><snm>Goldfarb</snm><fnm>I.</fnm></au>
    <au><snm>Banks Sills</snm><fnm>L.</fnm></au>
    <au><snm>Eliasi</snm><fnm>R.</fnm></au>
  </aug>
  <source>Phys. Rev. Lett.</source>
  <pubdate>2006</pubdate>
  <volume>97</volume>
  <fpage>206101</fpage>
</bibl>

<bibl id="B10">
  <title><p>{Effect of strain on the appearance of subcritical nuclei of Ge
  nanohuts on Si(001)}</p></title>
  <aug>
    <au><snm>Goldfarb</snm><fnm>I.</fnm></au>
  </aug>
  <source>Phys. Rev. Lett.</source>
  <pubdate>2005</pubdate>
  <volume>95</volume>
  <fpage>025501</fpage>
</bibl>

<bibl id="B11">
  <title><p>{Critical role of the surface reconstruction in the thermodynamic
  stability of $\{105\}$ Ge pyramids on Si(001)}</p></title>
  <aug>
    <au><snm>Raiteri</snm><fnm>P.</fnm></au>
    <au><snm>Migas</snm><fnm>D. B.</fnm></au>
    <au><snm>Miglio</snm><fnm>L</fnm></au>
    <au><snm>Rastelli</snm><fnm>A.</fnm></au>
    <au><snm>K{\"{a}}nel</snm><fnm>H.</fnm></au>
  </aug>
  <source>Phys. Rev. Lett.</source>
  <pubdate>2002</pubdate>
  <volume>88</volume>
  <issue>25</issue>
  <fpage>256103</fpage>
</bibl>

<bibl id="B12">
  <title><p>{Annealing of Ge nanocrystals on Si(001) at 550$^{\circ}$C:
  Metastability of huts and the stability of pyramids and domes}</p></title>
  <aug>
    <au><snm>Medeiros Ribeiro</snm><fnm>G.</fnm></au>
    <au><snm>Kamins</snm><fnm>T. I.</fnm></au>
    <au><snm>Ohlberg</snm><fnm>D. A. A.</fnm></au>
    <au><snm>Williams</snm><fnm>R. S.</fnm></au>
  </aug>
  <source>Phys. Rev. B</source>
  <pubdate>1998</pubdate>
  <volume>58</volume>
  <issue>7</issue>
  <fpage>3533</fpage>
</bibl>

<bibl id="B13">
  <title><p>{Spontaneous ordering of nanostructures on crystal
  surfaces}</p></title>
  <aug>
    <au><snm>Schukin</snm><fnm>V. A.</fnm></au>
    <au><snm>Bimberg</snm><fnm>D.</fnm></au>
  </aug>
  <source>Rev. Modern Phys.</source>
  <pubdate>1999</pubdate>
  <volume>71</volume>
  <issue>4</issue>
  <fpage>1125</fpage>
</bibl>

<bibl id="B14">
  <title><p>{Classification of Ge hut clusters in arrays formed by molecular
  beam epitaxy at low temperatures on the Si(001) surface}</p></title>
  <aug>
    <au><snm>Arapkina</snm><fnm>L. V.</fnm></au>
    <au><snm>Yuryev</snm><fnm>V. A.</fnm></au>
  </aug>
  <source>Phys. Usp.</source>
  <pubdate>2010</pubdate>
  <volume>53</volume>
  <issue>3</issue>
  <fpage>279</fpage>
  <note>arXiv:0907.4770</note>
</bibl>

<bibl id="B15">
  <title><p>{Ge/Si heterostructures with dense chains of stacked quantum
  dots}</p></title>
  <aug>
    <au><snm>Yuryev</snm><fnm>V. A.</fnm></au>
    <au><snm>Arapkina</snm><fnm>L. V.</fnm></au>
    <au><snm>Storozhevykh</snm><fnm>M. S.</fnm></au>
    <au><snm>Uvarov</snm><fnm>O. V.</fnm></au>
    <au><snm>Kalinushkin</snm><fnm>V. P.</fnm></au>
  </aug>
  <source>J. Nanoelectron. Optoelectron.</source>
  <pubdate>2014</pubdate>
  <volume>9</volume>
  <issue>2</issue>
  <fpage>196</fpage>
  <lpage>-218</lpage>
</bibl>

<bibl id="B16">
  <title><p>{An initial phase of Ge hut array formation at low temperature on
  Si(001)}</p></title>
  <aug>
    <au><snm>Arapkina</snm><fnm>L. V.</fnm></au>
    <au><snm>Yuryev</snm><fnm>V. A.</fnm></au>
  </aug>
  <source>J. Appl. Phys.</source>
  <pubdate>2011</pubdate>
  <volume>109</volume>
  <issue>10</issue>
  <fpage>104319</fpage>
  <note>arXiv:1009.3831</note>
</bibl>

<bibl id="B17">
  <title><p>{CMOS compatible dense arrays of Ge quantum dots on the Si(001)
  surface: Hut cluster nucleation, atomic structure, and array life cycle
  during UHV MBE growth}</p></title>
  <aug>
    <au><snm>Arapkina</snm><fnm>L. V.</fnm></au>
    <au><snm>Yuryev</snm><fnm>V. A.</fnm></au>
  </aug>
  <source>Nanoscale Res. Lett.</source>
  <pubdate>2011</pubdate>
  <volume>6</volume>
  <fpage>345</fpage>
  <note>arXiv:1009.3831</note>
</bibl>

<bibl id="B18">
  <title><p>{Ge quantum dot arrays grown by ultrahigh vacuum molecular-beam
  epitaxy on the Si(001) surface: nucleation, morphology, and CMOS
  compatibility}</p></title>
  <aug>
    <au><snm>Yuryev</snm><fnm>V. A.</fnm></au>
    <au><snm>Arapkina</snm><fnm>L. V.</fnm></au>
  </aug>
  <source>Nanoscale Res. Lett.</source>
  <pubdate>2011</pubdate>
  <volume>6</volume>
  <fpage>522</fpage>
  <note>arXiv:1104.2848</note>
</bibl>

<bibl id="B19">
  <title><p>{Ultrahigh vacuum scanning funnelling microscope
  GPI-300}</p></title>
  <aug>
    <au><snm>Eltsov</snm><fnm>K. N.</fnm></au>
    <au><snm>Klimov</snm><fnm>A. N.</fnm></au>
    <au><snm>Kosyakov</snm><fnm>A. N.</fnm></au>
    <au><snm>Obyedkov</snm><fnm>O. V.</fnm></au>
    <au><snm>Shevlyuga</snm><fnm>V. M.</fnm></au>
    <au><snm>Yurov</snm><fnm>VY</fnm></au>
  </aug>
  <source>Chemical state and atomic structure of fcc metal surfaces in chemical
  reaction with halogens</source>
  <publisher>Moscow, Russia: Nauka</publisher>
  <editor>V. I. Konov and K. N. Eltsov</editor>
  <series><title><p>Proc. of General Physics Institute</p></title></series>
  <pubdate>2003</pubdate>
  <volume>59</volume>
  <fpage>45</fpage>
</bibl>

<bibl id="B20">
  <title><p>{Structure and peculiarities of the $(8\times n)$-type Si(001)
  surface prepared in a molecular beam epitaxy chamber: A scanning tunneling
  microscopy study}</p></title>
  <aug>
    <au><snm>Arapkina</snm><fnm>L. V.</fnm></au>
    <au><snm>Shevlyuga</snm><fnm>V. M.</fnm></au>
    <au><snm>Yuryev</snm><fnm>V. A.</fnm></au>
  </aug>
  <source>JETP Lett.</source>
  <pubdate>2008</pubdate>
  <volume>87</volume>
  <fpage>215</fpage>
  <note>arXiv:0908.1346</note>
</bibl>

<bibl id="B21">
  <title><p>{Phase transition on the Si(001) clean surface prepared in UHV MBE
  chamber: A study by high resolution STM and in situ RHEED}</p></title>
  <aug>
    <au><snm>Arapkina</snm><fnm>L. V.</fnm></au>
    <au><snm>Yuryev</snm><fnm>V. A.</fnm></au>
    <au><snm>Chizh</snm><fnm>K. V.</fnm></au>
    <au><snm>Shevlyuga</snm><fnm>V. M.</fnm></au>
    <au><snm>Storojevyh</snm><fnm>M. S.</fnm></au>
    <au><snm>Krylova</snm><fnm>L. A.</fnm></au>
  </aug>
  <source>Nanoscale Res. Lett.</source>
  <pubdate>2011</pubdate>
  <volume>6</volume>
  <fpage>218</fpage>
  <note>arXiv:1009.3909</note>
</bibl>

<bibl id="B22">
  <title><p>{Phase transition between ${(2\times 1)}$ and ${c(8\times 8)}$
  reconstructions observed on the Si(001) surface around
  600$^{\circ}$C}</p></title>
  <aug>
    <au><snm>Arapkina</snm><fnm>L. V.</fnm></au>
    <au><snm>Yuryev</snm><fnm>V. A.</fnm></au>
    <au><snm>Shevlyuga</snm><fnm>V. M.</fnm></au>
    <au><snm>Chizh</snm><fnm>K. V.</fnm></au>
  </aug>
  <source>JETP Lett.</source>
  <pubdate>2010</pubdate>
  <volume>92</volume>
  <issue>5</issue>
  <fpage>310</fpage>
  <note>arXiv:1007.0329</note>
</bibl>

<bibl id="B23">
  <title><p>{Application of hydrogenation to low-temperature cleaning of the
  Si(001) surface in the processes of molecular-beam epitaxy: Investigation by
  scanning tunneling microscopy, reflected high-energy electron diffraction,
  and high resolution transmission electron microscopy}</p></title>
  <aug>
    <au><snm>Arapkina</snm><fnm>L. V.</fnm></au>
    <au><snm>Krylova</snm><fnm>L. A.</fnm></au>
    <au><snm>Chizh</snm><fnm>K. V.</fnm></au>
    <au><snm>Chapnin</snm><fnm>V. A.</fnm></au>
    <au><snm>Uvarov</snm><fnm>O. V.</fnm></au>
    <au><snm>Yuryev</snm><fnm>V. A.</fnm></au>
  </aug>
  <source>J. Appl. Phys.</source>
  <pubdate>2012</pubdate>
  <volume>112</volume>
  <issue>1</issue>
  <fpage>014311</fpage>
  <note>arXiv:1202.6224</note>
</bibl>

<bibl id="B24">
  <title><p>{WSxM: A software for scanning probe microscopy and a tool for
  nanotechnology}</p></title>
  <aug>
    <au><snm>Horcas</snm><fnm>I.</fnm></au>
    <au><snm>Fernandez</snm><fnm>R.</fnm></au>
    <au><snm>Gomez Rodriguez</snm><fnm>J. M.</fnm></au>
    <au><snm>Colchero</snm><fnm>J.</fnm></au>
    <au><snm>Gomez Herrero</snm><fnm>J.</fnm></au>
    <au><snm>Baro</snm><fnm>A. M.</fnm></au>
  </aug>
  <source>Rev. Sci. Instrum.</source>
  <pubdate>2007</pubdate>
  <volume>78</volume>
  <fpage>013705</fpage>
</bibl>

<bibl id="B25">
  <title><p>{Dense chains of stacked quantum dots in Ge/Si
  heterostructures}</p></title>
  <aug>
    <au><snm>Yuryev</snm><fnm>V. A.</fnm></au>
    <au><snm>Arapkina</snm><fnm>L. V.</fnm></au>
    <au><snm>Storozhevykh</snm><fnm>M. S.</fnm></au>
    <au><snm>Uvarov</snm><fnm>O. V.</fnm></au>
    <au><snm>Kalinushkin</snm><fnm>V. P.</fnm></au>
  </aug>
  <source>Nanotechnology VI</source>
  <publisher>SPIE, Bellingham, WA</publisher>
  <editor>Riner Adelung</editor>
  <series><title><p>Proc. SPIE</p></title></series>
  <pubdate>2013</pubdate>
  <volume>8766</volume>
  <fpage>876606</fpage>
  <note>{SPIE Paper Number 8766-5}</note>
</bibl>

<bibl id="B26">
  <title><p>{Reversal of step roughness on Ge-covered vicinal
  Si(001)}</p></title>
  <aug>
    <au><snm>Wu</snm><fnm>F</fnm></au>
    <au><snm>Chen</snm><fnm>X</fnm></au>
    <au><snm>Zhang</snm><fnm>Z</fnm></au>
    <au><snm>Lagally</snm><fnm>M. G.</fnm></au>
  </aug>
  <source>Phys. Rev. Lett.</source>
  <pubdate>1995</pubdate>
  <volume>74</volume>
  <fpage>574</fpage>
</bibl>

<bibl id="B27">
  <title><p>{STM study of initial stage of Ge epitaxy on Si(001)}</p></title>
  <aug>
    <au><snm>Iwawaki</snm><fnm>F.</fnm></au>
    <au><snm>Tomitori</snm><fnm>M.</fnm></au>
    <au><snm>Nishikawa</snm><fnm>O.</fnm></au>
  </aug>
  <source>Ultramicroscopy</source>
  <pubdate>1992</pubdate>
  <volume>42--44</volume>
  <fpage>902</fpage>
</bibl>

<bibl id="B28">
  <title><p>{Evolution of the strain relaxation in a Ge layer on Si(001) by
  reconstruction and intermixing}</p></title>
  <aug>
    <au><snm>Voigtl{\"{a}}nder</snm><fnm>B.</fnm></au>
    <au><snm>K{\"{a}}stner</snm><fnm>M.</fnm></au>
  </aug>
  <source>Phys. Rev. B</source>
  <pubdate>1999</pubdate>
  <volume>60</volume>
  <fpage>R5121</fpage>
</bibl>

<bibl id="B29">
  <title><p>{Evolution of Ge/Si(001) wetting layer during Si overgrowth and
  crossover between thermodynamic and kinetic behavior}</p></title>
  <aug>
    <au><snm>Migas</snm><fnm>D. B.</fnm></au>
    <au><snm>Raiteri</snm><fnm>P.</fnm></au>
    <au><snm>Miglio</snm><fnm>L.</fnm></au>
    <au><snm>Rastelli</snm><fnm>A.</fnm></au>
    <au><snm>K{\"{a}}nel</snm><fnm>H.</fnm></au>
  </aug>
  <source>Phys. Rev. B</source>
  <pubdate>2004</pubdate>
  <volume>69</volume>
  <fpage>235318</fpage>
</bibl>

<bibl id="B30">
  <title><p>{Pathway to the strain-driven two-dimensional to three-dimensional
  transitions during growth of Ge on Si(001)}</p></title>
  <aug>
    <au><snm>Vailionis</snm><fnm>A.</fnm></au>
    <au><snm>Cho</snm><fnm>B.</fnm></au>
    <au><snm>Glass</snm><fnm>G.</fnm></au>
    <au><snm>Desjardins</snm><fnm>P.</fnm></au>
    <au><snm>Cahill</snm><fnm>DG</fnm></au>
    <au><snm>Greene</snm><fnm>J. E.</fnm></au>
  </aug>
  <source>Phys. Rev. Lett.</source>
  <pubdate>2000</pubdate>
  <volume>85</volume>
  <fpage>3672</fpage>
  <lpage>-3675</lpage>
</bibl>

<bibl id="B31">
  <title><p>{On atomic structure of Ge huts growing on the Ge/Si(001) wetting
  layer}</p></title>
  <aug>
    <au><snm>Arapkina</snm><fnm>LV</fnm></au>
    <au><snm>Yuryev</snm><fnm>VA</fnm></au>
  </aug>
  <source>J. Appl. Phys.</source>
  <pubdate>2013</pubdate>
  <volume>114</volume>
  <fpage>104304</fpage>
  <note>arXiv:1210.2974</note>
</bibl>

<bibl id="B32">
  <title><p>{Nucleation of Ge clusters at high temperatures on Ge/Si(001)
  wetting layer}</p></title>
  <aug>
    <au><snm>Yuryev</snm><fnm>V. A.</fnm></au>
    <au><snm>Arapkina</snm><fnm>L. V.</fnm></au>
  </aug>
  <source>J. Appl. Phys.</source>
  <pubdate>2012</pubdate>
  <volume>111</volume>
  <fpage>094307</fpage>
  <note>arXiv:1105.6012</note>
</bibl>

<bibl id="B33">
  <title><p>{Ge/Si(001) heterostructures with dense arrays of Ge quantum dots:
  morphology, defects, photo-emf spectra and terahertz
  conductivity}</p></title>
  <aug>
    <au><snm>Yuryev</snm><fnm>V. A.</fnm></au>
    <au><snm>Arapkina</snm><fnm>L. V.</fnm></au>
    <au><snm>Storozhevykh</snm><fnm>M. S.</fnm></au>
    <au><snm>Chapnin</snm><fnm>V. A.</fnm></au>
    <au><snm>Chizh</snm><fnm>K. V.</fnm></au>
    <au><snm>Uvarov</snm><fnm>O. V.</fnm></au>
    <au><snm>Kalinushkin</snm><fnm>V. P.</fnm></au>
    <au><snm>Zhukova</snm><fnm>E. S.</fnm></au>
    <au><snm>Prokhorov</snm><fnm>A. S.</fnm></au>
    <au><snm>Spektor</snm><fnm>I. E.</fnm></au>
    <au><snm>Gorshunov</snm><fnm>B. P.</fnm></au>
  </aug>
  <source>Nanoscale Res. Lett.</source>
  <pubdate>2012</pubdate>
  <volume>7</volume>
  <fpage>414</fpage>
  <note>arXiv:1204.2509</note>
</bibl>

<bibl id="B34">
  <title><p>{Prepyramid-to-pyramid transition of SiGe islands on
  Si(001)}</p></title>
  <aug>
    <au><snm>Rastelli</snm><fnm>A.</fnm></au>
    <au><snm>K{\"{a}}nel</snm><fnm>H.</fnm></au>
    <au><snm>Spencer</snm><fnm>B. J.</fnm></au>
    <au><snm>Tersoff</snm><fnm>J.</fnm></au>
  </aug>
  <source>Phys. Rev. B</source>
  <pubdate>2003</pubdate>
  <volume>68</volume>
  <fpage>115301</fpage>
</bibl>

<bibl id="B35">
  <title><p>{Morphological evolution of Ge/Si(001) quantum dot rings formed at
  the rim of wet-etched pits}</p></title>
  <aug>
    <au><snm>Grydlik</snm><fnm>M</fnm></au>
    <au><snm>Brehm</snm><fnm>M</fnm></au>
    <au><snm>Sch{\"{a}}ffler</snm><fnm>F</fnm></au>
  </aug>
  <source>Nanoscale Res. Lett.</source>
  <pubdate>2012</pubdate>
  <volume>7</volume>
  <fpage>601</fpage>
</bibl>

<bibl id="B36">
  <title><p>{Die Grundlagen der Thermodynamik}</p></title>
  <aug>
    <au><snm>Ehrenfest Afanassjewa</snm><fnm>T</fnm></au>
  </aug>
  <publisher>Leiden, Germany: E.\,J. Brill</publisher>
  <pubdate>1956</pubdate>
</bibl>

<bibl id="B37">
  <title><p>{Delayed plastic relaxation limit in {SiGe} islands grown by {Ge}
  diffusion from a local source}</p></title>
  <aug>
    <au><snm>Vanacore</snm><fnm>G. M.</fnm></au>
    <au><snm>Nicotra</snm><fnm>G.</fnm></au>
    <au><snm>Zani</snm><fnm>M.</fnm></au>
    <au><snm>Bollani</snm><fnm>M.</fnm></au>
    <au><snm>E. Bonera</snm><fnm>FM</fnm></au>
    <au><snm>Capellini</snm><fnm>G.</fnm></au>
    <au><snm>Isella</snm><fnm>G.</fnm></au>
    <au><snm>Osmond</snm><fnm>J.</fnm></au>
    <au><snm>A. Picco</snm><fnm>FB</fnm></au>
    <au><snm>Tagliaferri</snm><fnm>A.</fnm></au>
  </aug>
  <source>J. Appl. Phys.</source>
  <pubdate>2015</pubdate>
  <volume>117</volume>
  <fpage>104309</fpage>
</bibl>

<bibl id="B38">
  <title><p>Monitoring the kinetic evolution of self-assembled {SiGe} islands
  grown by {Ge} surface thermal diffusion from a local source</p></title>
  <aug>
    <au><snm>Vanacore</snm><fnm>G. M.</fnm></au>
    <au><snm>Zani</snm><fnm>M.</fnm></au>
    <au><snm>Bollani</snm><fnm>M.</fnm></au>
    <au><snm>Bonera</snm><fnm>E.</fnm></au>
    <au><snm>Nicotra</snm><fnm>G.</fnm></au>
    <au><snm>Osmond</snm><fnm>J.</fnm></au>
    <au><snm>Capellini</snm><fnm>G</fnm></au>
    <au><snm>Isella</snm><fnm>G.</fnm></au>
    <au><snm>Tagliaferri</snm><fnm>A.</fnm></au>
  </aug>
  <source>Nanotechnology</source>
  <pubdate>2014</pubdate>
  <volume>25</volume>
  <fpage>135606</fpage>
</bibl>

<bibl id="B39">
  <title><p>{Ge/Si(001) wetting layer formed by Ge deposition at room
  temperature followed by annealing at 600{\,\textcelsius} and its structural
  features}</p></title>
  <aug>
    <au><snm>Storozhevykh</snm><fnm>MS</fnm></au>
    <au><snm>Arapkina</snm><fnm>LV</fnm></au>
    <au><snm>Yuryev</snm><fnm>VA</fnm></au>
  </aug>
  <source>Posted on arXiv:1409.5422v2</source>
  <pubdate>2014</pubdate>
</bibl>

</refgrp>
} 




\newpage
\section*{Figures}

  \begin{figure}[h!]
\begin{flushleft}
\includegraphics[scale=.145]{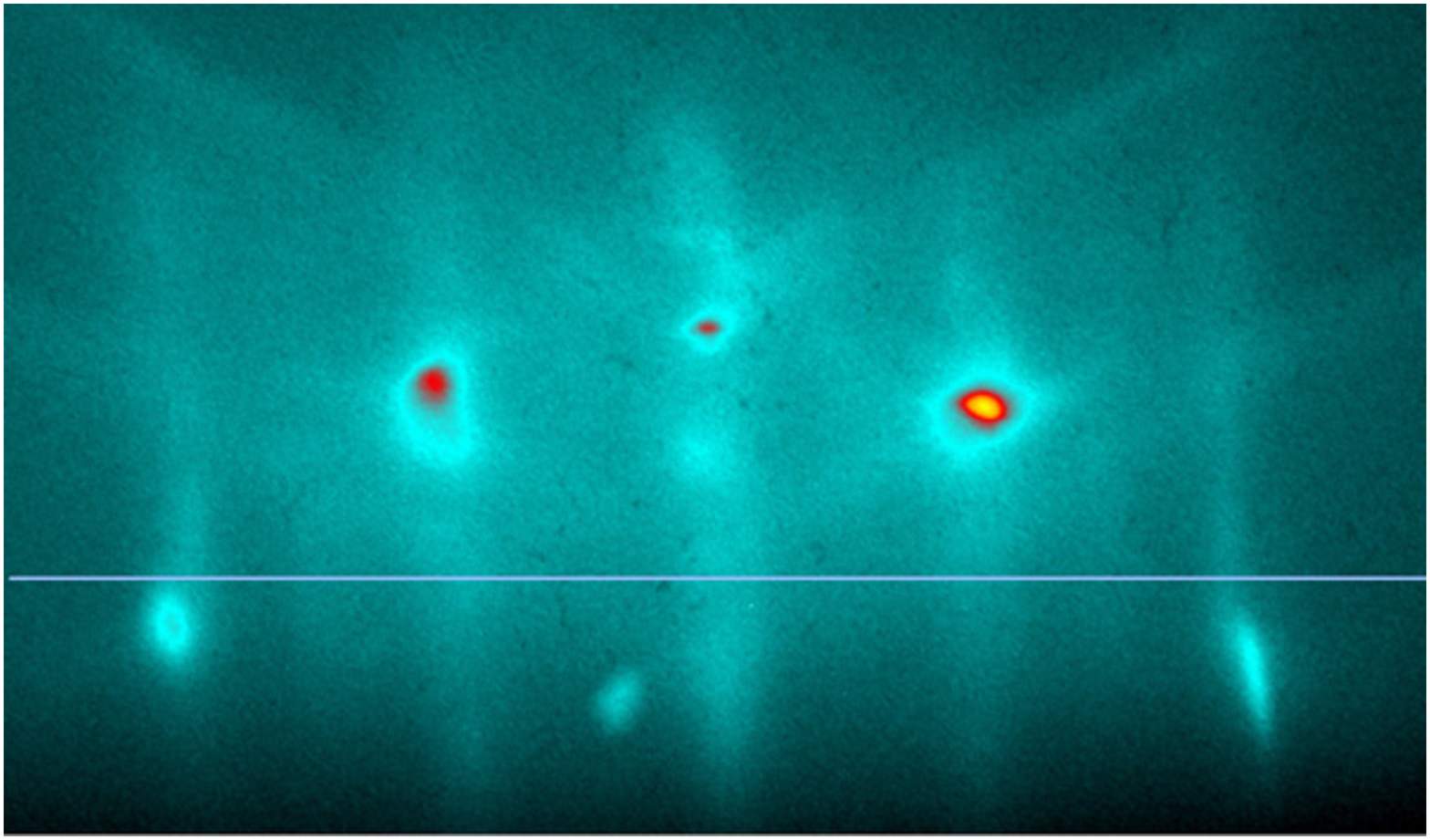}\,(a)
\includegraphics[scale=.15]{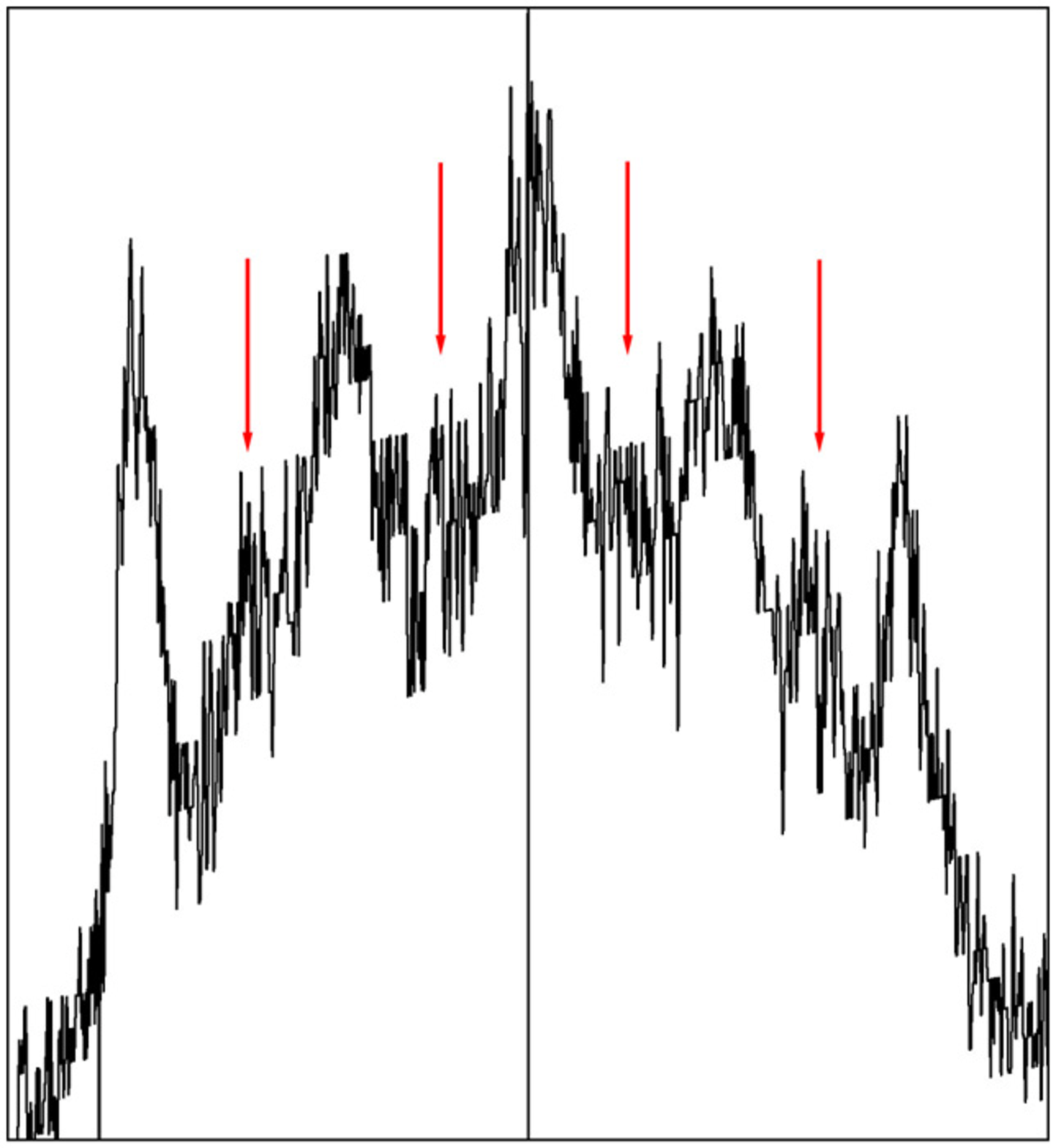}\,(b)\\~\\
\includegraphics[scale=1]{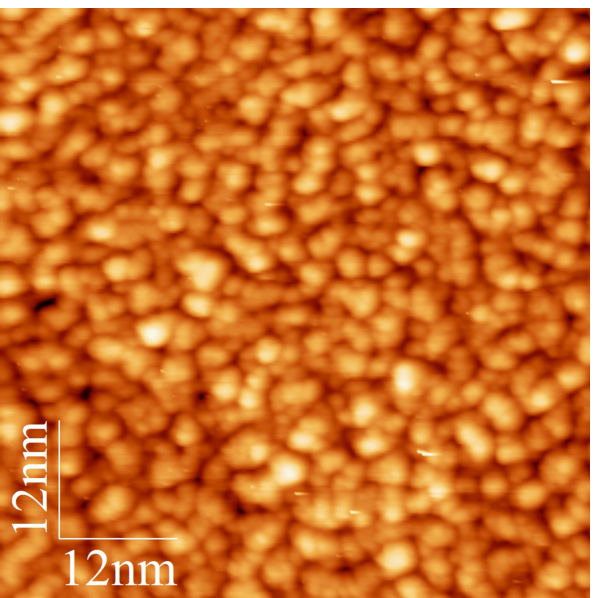}\,
\includegraphics[scale=.93]{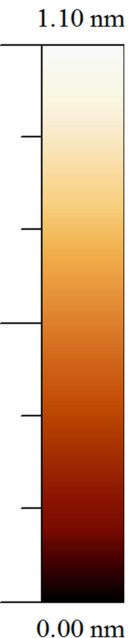}\,(c)
\end{flushleft}
  \caption{\csentence{\label{fig:RHEED_before}
A RHEED pattern and an STM image of the initial Ge film.
}
(a) The RHEED pattern obtained at the room temperature after Ge film deposition before sample heating
(the effective thickness {\em h}$_{\rm Ge}$ = 7\,{\AA};  [110] azimuth, {\em E} = 10 keV)
and 
(b)  its profile taken along the light line in the panel (a); arrows in the panel (b) indicate the weak {\textonehalf}-reflexes virtually unobservable at the fluorescent screen during the experiments
but visible in the panel (a)
which demonstrate that the 2\,$\times$\,1 structure occupies a minor part of the film surface area.
(c) The STM image  demonstrates a grainy disordered structure of the film.
}
      \end{figure}

\begin{figure}[h!]
\includegraphics[scale=.9]{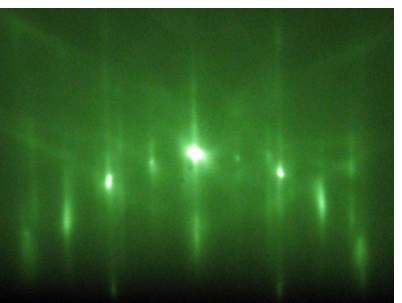}(a)
\includegraphics[scale=.9]{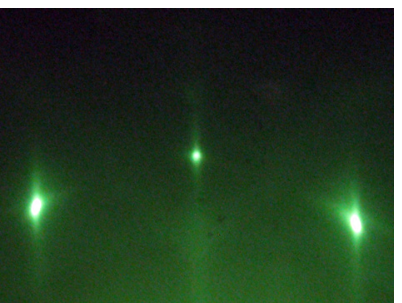}(b)
  \caption{\csentence{\label{fig:RHEED_after}
RHEED patterns  obtained at room temperature after sample cooling.
}
{\em E} = 10 keV;
(a) [110] and
(b) [100] azimuths;  
streaks of the 2\,$\times$\,1 structure and 3D reflexes (discontinuous streaks) are observed.
}
      \end{figure}

\begin{figure}[h!]
\includegraphics[scale=1]{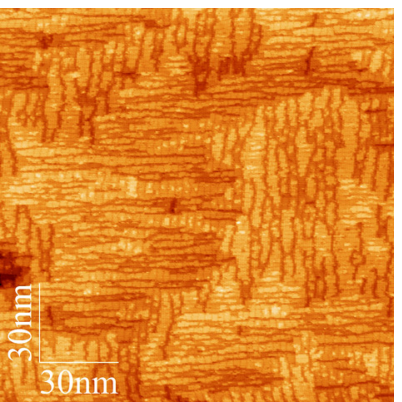}(a)
\includegraphics[scale=1]{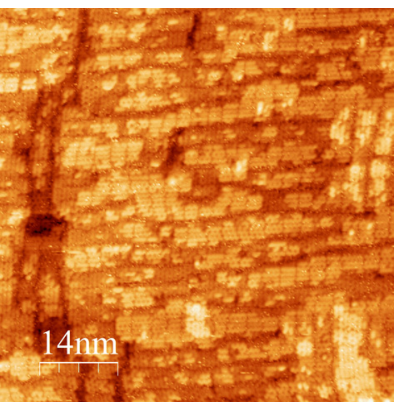}(b)
\includegraphics[scale=1]{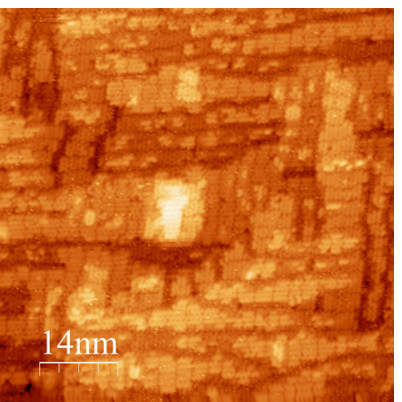}(c)
  \caption{\csentence{\label{fig:WL}
STM micrographs of the annealed sample obtained at different points on the surface.
}
A usual  {\em M\,}$\times$\,{\em N} reconstruction composed by 
{\em p}(2\,$\times$\,2)
and 
{\em c}(4\,$\times$\,2)
reconstructed patches is observed (the {\em p}(2\,$\times$\,2) reconstruction is seen as in-phase zigzags, the {\em c}(4\,$\times$\,2) one is seen as anti-phase zigzags).
The images (a) to (c) were obtained at different points on a sample.
}
      \end{figure}

\begin{figure}[h!]
\includegraphics[scale=1]{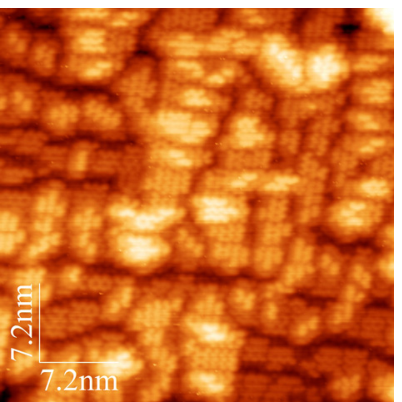}(a)
\includegraphics[scale=1]{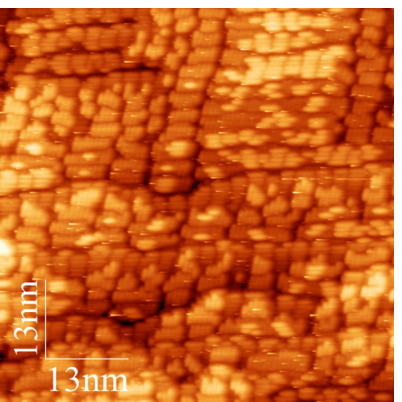}(b)
\includegraphics[scale=1]{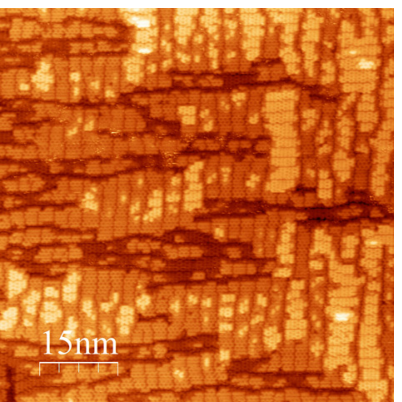}(c)
  \caption{\csentence{\label{fig:WL-360_600_650}
STM images of the Ge/Si(001) wetting layers grown at different conditions.
}
(a) {\em T}$_{\rm gr}$ = 360{\,\textcelsius}, {\em h}$_{\rm Ge}$ = 4\,{\AA};
(b) {\em T}$_{\rm gr}$ = 600{\,\textcelsius}, {\em h}$_{\rm Ge}$ = 6\,{\AA};
(c) {\em T}$_{\rm gr}$ = 650{\,\textcelsius}, {\em h}$_{\rm Ge}$ = 4\,{\AA}.
}
      \end{figure}

\begin{figure}[h!]
\includegraphics[scale=.9]{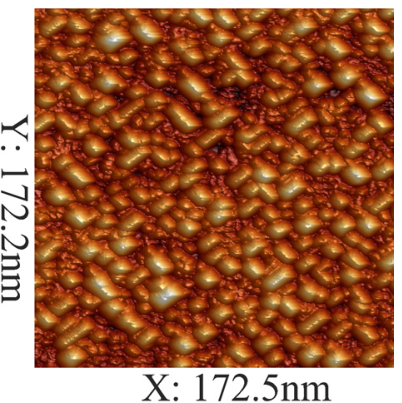}(a)
\includegraphics[scale=.9]{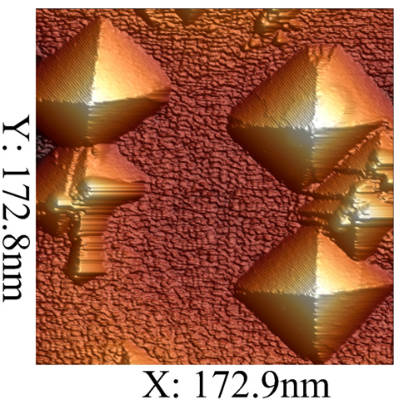}(b)
  \caption{\csentence{\label{fig:array-600_360C}
STM micrographs of Ge huts forming at different conditions.
}
(a) {\em T}$_{\rm gr}$ = 360{\,\textcelsius}, {\em h}$_{\rm Ge}$\,=  7\,{\AA};
(b){\em T}$_{\rm gr}$ = 600{\,\textcelsius}, {\em h}$_{\rm Ge}$ =  6\,{\AA}.
}
      \end{figure}

\begin{figure}[h!]
\includegraphics[scale=.86]{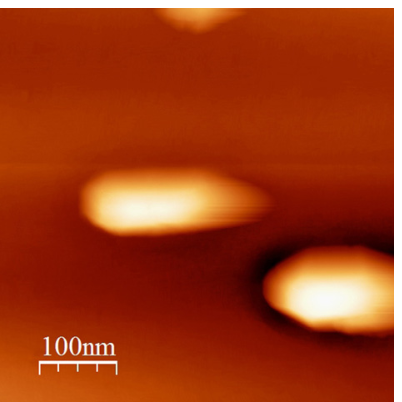}(a)
\includegraphics[scale=.85]{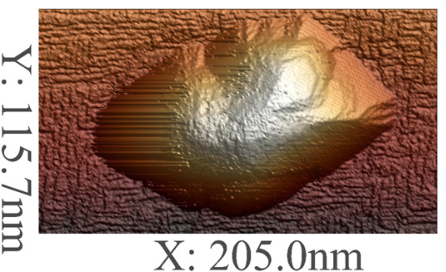}(b)
\includegraphics[scale=.85]{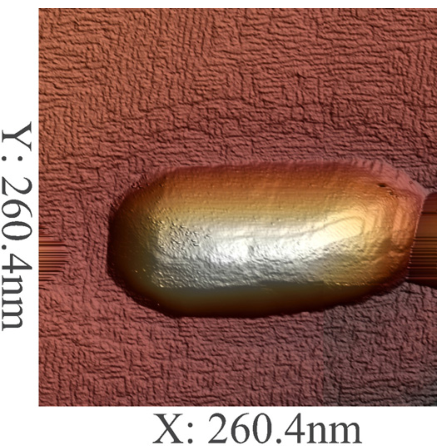}(c)
\includegraphics[scale=.85]{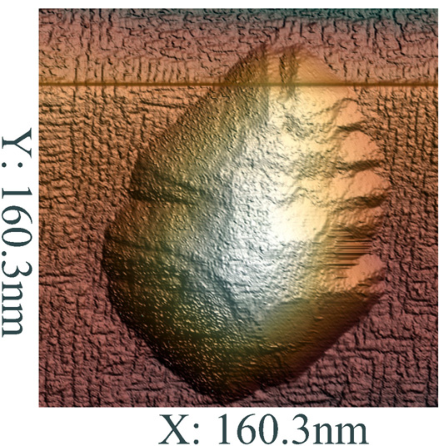}(d)
  \caption{\csentence{\label{fig:clusters}
STM images of Ge clusters observed on the wetting layer after annealing at 600\,{\textcelsius} of the Ge film deposited at the room temperature.
}
{\em h}$_{\rm Ge}$\,=  7\,{\AA}.
}
      \end{figure}

\end{backmatter}
\end{document}